\documentclass[aps,prd,twocolumn,groupedaddress,nofootinbib]{revtex4-1}
\usepackage{tensor}
\usepackage{amsmath}
\usepackage{amssymb}
\usepackage{mathtools}
\usepackage{graphicx}
\usepackage{wrapfig}
\usepackage{float}
\usepackage[
	colorlinks=true,
	linkcolor=blue,
	citecolor=blue,
	filecolor=magenta,
	urlcolor=blue
]{hyperref}

\begin{document}
\title{Response to Comment on ``Reexamining $f(R,T)$ Gravity''}
\author{Sarah B. Fisher}
\author{Eric D. Carlson}
\email{ecarlson@wfu.edu}
\affiliation{Department of Physics, Wake Forest University, 1834 Wake Forest Road, Winston-Salem, North Carolina 27109, USA}
\date{\today}

\begin{abstract}
    Harko and Moraes \cite{Harko:2020ivb} claim that in $f(R,T)$ gravity with $f(R,T)=f_1(R)+f_2(T)$, the term $f_2(T)$ cannot be incorporated in the matter Lagrangian ${\cal L}_m$. A careful examination of their Comment finds that they have made several dubious assumptions without indicating any errors in our work. Most notably, they have unjustifiably claimed that the two terms ${\cal L}_m$ and $f_2(T)$ are of ``different origin,'' and their inference that only the first contributes to the energy momentum tensor is arbitrary.  Also significant, their derivation of equations of motion from a Lagrangian formulation, imposing conservation constraints {\it ad hoc} rather than via Lagrange multipliers, leads to inconsistent conclusions.
\end{abstract}

\maketitle

Harko and Moraes \cite{Harko:2020ivb} have disputed our claim that in $f(R,T)$ gravity, when $f(R,T)=f_1(R)+f_2(T)$, it is generally possible to incorporate $f_2(T)$ into ${\cal L}_m$.  In section II, they claim that they can come up with examples where the term $f_2(T)$ cannot be incorporated into ${\cal L}_m$ in a ``natural'' way. In section III, they claim that we have mistakenly claimed that quantities such as $\rho$ and $n$ do not correspond to the physical energy density and number density.  We will address these two issues in turn.

First, they argue that for a free scalar field, there are functions $f_2(T)$ such that the Lagrangian density ${\cal L}' = {\cal L}_m + \frac12\kappa^{-2}f_2(T)$ cannot necessarily be written in the ``natural'' form $\frac12 \partial_\mu \Phi \partial^\mu \Phi - U(\Phi)$.  We agree with this statment.  However, when we stated that $f_2(T)$ can be included in ${\cal L}_m$, we only meant that the resulting Lagrangian could be written in the form  ${\cal L}'_m ={\cal L}'_m(\phi,\partial_\mu \phi)$.  Their only complaint seems to be that somehow this Lagrangian is unnatural.  But there is no reason to restrict consideration to interactions of the form $U(\Phi)$.  Derivative couplings occur in a variety of contexts, such as non-Abelian gauge theories, effective theories, and general relativity itself, and indeed, it is unnatural to demand that the Lagrangian must take the form $\frac12 \partial_\mu \Phi \partial^\mu \Phi - U(\Phi)$.  If this clarification helps our readers understand our meaning, then we welcome this Comment. 

The remainder of the Comment is focused on our work with a perfect fluid.  In this case, the authors attempt to define physical quantities as ``quantities that are obtained from the microscopic distribution function of the particles.''  Leaving aside the philosophical question of whether it is appropriate to use the word ``physical'' to reference an essentially mathematical property, we find an admission in the Comment that the calculation in question has not been performed in $f(R,T)$ gravity.  Their subsequent assumption that $f_2(T)$ should not be incorporated in this microscopic distribution function (and therefore remains outside the matter Lagrangian) is unjustified.  In particular, we demonstrated that it {\it must} be incorporated for a linear interaction $f_2(T) \propto T$ for a free scalar field. A similar effect occurs for fermions.

As an example, they state that for a degenerate relativistic fermionic gas, $p \propto \rho^{4/3}$.   It is true that a non-interacting degenerate relativistic fermionic gas will have its physical pressure $p = \frac13\rho \propto n^{4/3}$, but the applicability of this approximation will depend on the mass and the amount of interaction, which both depend not only on ${\cal L}_m$, but also on $f_2(T)$.  There is simply no way to ignore $f_2(T)$ when considering whether a given fluid is describable as a non-interactimg degenerate relativistic fermionic gas.

The authors attempt to sidestep this argument by claiming that the term $f_2(T)$ is ``of different origins,'' but they do not justify this artificial separation of ${\cal L}_m$ and $f_2(T)$.  These two terms depend on the same variables and therefore simply cannot be divided this way.  By stating that they are of different origins and cannot be combined, the authors ignore the central point of our paper.

Then they state that we concluded, because the quantities derived from the combined effects of ${\cal L}_m$ and $f_2(T)$ are conserved, that these are the true density, pressure, and particle number.  They have our argument backwards.  In fact, we note that because ${\cal L}_m$ and $f_2(T)$ are added and are both functions of the same variables, there is no way to physically disentangle their effects, and therefore any physical measurment of pressure, etc.\ will see the effects of both.  We then demonstrated that when both are included in ${\cal L}_m^\prime$, there is a modified conserved current $J_\mu^\prime$ and conserved energy-momentum tensor $T_{\mu\nu}^\prime$.  To ignore $f_2(T)$ while including ${\cal L}_m$ is no more justified than if one were to divide ${\cal L}_m$ into two terms, and include one and exclude the other.

The authors then go into a calculation attempting to demonstrate how one can ``correctly'' find the stress-energy tensor, etc., by starting with a simple Lagrangian ${\cal L}_m = -\rho$ and then considering only variations of various quantities that do not violate things like conservation of comoving particle number and comoving entropy per particle.   They seem to feel that somehow our approach is inferior, in that we find these conservation laws coming from our equations of motion, rather than being simply assumed.  In fact, conservation laws {\it should} come about because of the equations of motion.  In situations where we wish to impose conservation laws on otherwise unconstrained variables, the standard procedure is to use Lagrange multipliers.  This is the whole purpose of Lagrange multipliers, and this is exactly what we have done.

If one imposes conservation laws externally, without including such conservation-imposing terms, one can easily end up with inconsistent equations, and indeed, this is what has happened in this case.  To demonstrate, let us note that this Comment, though never explicitly writing down the energy-momentum tensor, has expressions like $T=\rho-3p$, which implies that they believe it takes the standard perfect fluid form.  We also note that, as pointed out in \cite{Harko:2011kv}, $\nabla_\mu T^{\mu\nu} \ne 0$.  We believe, therefore, that the authors accept all of the following equations to be true:
\begin{subequations}\begin{eqnarray}
T^{\mu\nu} &=& (\rho + p)u^\mu u^\nu - pg^{\mu\nu} \; , \label{Standard_Energy_Momentum} \\
\nabla_\mu \left( n u^\mu\right) &=& 0 \label{Conserved_Particle_Number}\; ,\\
n \left( \frac{\partial \rho}{\partial n}\right)_s &=& \rho + p \; , \label{Pressure_Expression} \\
u^\mu \partial_\mu s &=& 0 \; , \label{Conserved_Entropy}\\
\nabla_\mu T^{\mu\nu} &\ne& 0 \; . \label{Non_Conserved_Tensor}
\end{eqnarray}\end{subequations}
However, using Eqs.~(\ref{Standard_Energy_Momentum}-\ref{Conserved_Entropy}) (together with $u_\mu u^\mu = 1$) one can show that $u_\nu \nabla_\mu T^{\mu\nu} = 0$, which is inconsistent with Eq.~(\ref{Non_Conserved_Tensor}).  This demonstrates convincingly that the Comment's assumption of conservation laws without incorporating the appropriate Lagrange multipliers can lead to inconsistent and misleading equations.

In summary, the authors presuppose the following: derivative couplings are unnatural; $f_2(T)$ can be neglected when calculating distribution functions;  $f_2(T)$ is ``of different origins'' than ${\cal L}_m$; the quantities calculated from the distribution functions are the same ones one would measure empirically; and the bare particle number is conserved, without reference to the equations of motion.  Most of these assumptions are hidden, all are questionable, and some are precisely what we demonstrated to be false.

Nowhere do the authors explain where we are supposed to have gone wrong in order to reach false conclusions.  Rather, from the aforementioned assumptions, they draw their own conclusions, which contradict ours.  They claim this raises concerns about the validity of our results.  In fact, such a proof from dubious premises, reaching conclusions which contradict published results, with no explanation of where said research might have gone wrong, strongly suggests that those premises are, in fact, false.

We stand by our claim that $f_2(T)$ can and generally should be incorporated into ${\cal L}_m$, and that treating ${\cal L}_m$ and $f_2(T)$ as terms with ``different origins'' is inconsistent and has no physical meaning.


%

\end{document}